\documentclass[a4paper,11pt]{article}
\pdfoutput=1 

\usepackage{jcappub} 

\usepackage[T1]{fontenc} 
\usepackage{comment}
\usepackage{xcolor}
\usepackage{amsmath}
\title{\boldmath Consequences of neutron decay inside neutron stars}

\author[a,1]{Wasif Husain,\note{Corresponding author.}}
\author[a,b]{Theo F. Motta}
\author[a]{Anthony W. Thomas}

\affiliation[a]{ARC Centre of Excellence for Dark Matter Particle Physics and CSSM, Department of
Physics, University of Adelaide, SA 5005, Australia}
\affiliation[b]{Institut f\"{u}r Theoretische Physik, Justus-Liebig-Universit\"{a}t Giessen, 35392 Giessen, Germany}

\emailAdd{wasif.husain@adelaide.edu.au}
\emailAdd{theo.ferraz-motta@theo.physik.uni-giessen.de}
\emailAdd{anthony.thomas@adelaide.edu.au}

\abstract{The hypothesis that neutrons might decay into dark matter is explored using neutron stars as a testing ground. It is found that in order to obtain stars with masses at the upper end of those observed, the dark matter must experience a relatively strong self-interaction. Conservation of baryon number and energy then require that the star must undergo some heating, with a decrease in radius, leading to an increase in speed of rotation over a period of days.}

\begin{document}
\maketitle
\flushbottom

\section{Introduction}
For some years we have been confronted with a puzzling  discrepancy between the two different methods of measuring the neutron lifetime~\cite{RevModPhys.83.1173}. In particular, neutrons observed to decay into protons in a 
beam~\cite{PhysRevLett.111.222501,Otono:2016fsv,Olive_2016} yield a lifetime roughly one percent longer than those trapped in a bottle, whose decay mode is undetermined~\cite{Serebrov:2017bzo,Pattie:2017vsj,TAN2019134921,PhysRevC.85.065503}. This controversy inspired a remarkably precise measurement of the lifetime of neutrons trapped in a bottle by the UCN$\tau$ Collaboration~\cite{UCNt:2021pcg}, with the result $877.75 \pm 0.28_{\rm stat} + 0.22 - 0.16_{\rm syst}$ s. This agrees well with other recent measurements such as  Ref.~\cite{doi:10.1126/science.aan8895}, which reported a neutron lifetime  $877.7 \pm 0.7$ s  and Ref.~\cite{PhysRevC.97.055503}, which reported a value $881.5 \pm 0.7$ s.  For comparison, the measurements using a beam of neutrons yield a lifetime of approximately 887.7 s \cite{doi:10.1126/science.aan8895}. Thus the average lifetime values determined by the two different techniques differ by approximately 4$\sigma$ (around 8 seconds). 

A possible explanation for this discrepancy, which has excited a great deal of interest, was proposed by Fornal and Grinstein ~\cite{Grinstein:2018ptl}. The idea is that in addition to the standard decay mode, $n \, \rightarrow \, p \, e^- \, \bar{\nu_e}$ observed in the beam decay experiments,  neutrons might have a small branching ratio for decay into dark sector particles (dark matter). The idea is that the dark matter particles would remain undetected in the beam experiments because of their very weak interactions with the Standard Model particles, whereas in the bottle experiments they will contribute to the total decay rate of neutrons.
This proposal is an alternative to the hypothesis that a neutron might oscillate into its mirror counterpart, which 
had previously been
ruled out~\cite{Serebrov:2007gw}. Some of the other decay channels are discussed in Ref.~\cite{Ivanov:2018vit}

The suggestion of Fornal and Grinstein requires the existence of a dark decay channel, including a dark fermion, $\chi$,  which is very nearly degenerate with the neutron
\begin{equation}
    n \xrightarrow \qquad \chi + \phi \, .
    \label{eq:decay}
\end{equation}
In fact, a narrow range of mass of dark fermions may be allowed by the systematic of stable nuclei 937.9 MeV < $m_\chi$ < 938.7 MeV~\cite{Grinstein:2018ptl}. That the accompanying boson, $\phi$, cannot be a photon was established very quickly~\cite{Tang:2018eln}.
Serebrov {\em et al.}~\cite{Serebrov:2018mva} argued that this proposal might also solve the experimental inconsistency of the reactor anti-neutrino anomaly. 

That it should be possible to use the properties of neutron stars to test this hypothesis. Indeed, if they are non-interacting, this decay mode reduces the maximum possible mass of a neutron star to around $0.7 \, M_\odot$~\cite{Motta:2018rxp,PhysRevLett.121.061801,McKeen:2018xwc}, well below the masses of known stars, which range up to $2 \, M_\odot$~\cite{Ozel:2016oaf,2010Natur.467.1081D,2013Sci...340..448A}.  The decay process shown in Eq.~(1) would allow a fraction of the neutrons to decay into DM particles, beginning right after the birth of the neutron star. The DM particles will be in chemical equilibrium with the normal nuclear matter and would remain in the gravitational potential well of the neutron star.  The basic physics is that, if the dark fermions are non-interacting the conversion of neutrons, with their high chemical potential, into dark matter particles is highly energetically favored.

Thus, if the Fornal-Grinstein idea is to survive, the dark matter fermions must experience a repulsive self-interaction. Then, using the observational constraints on neutron star properties, it is possible that we can deduce constraints on the properties of this dark matter~\cite{Motta:2018bil,Grinstein:2018ptl,PhysRevLett.121.061801,Cline:2018ami,Berryman:2022zic,Strumia:2021ybk,2021_ramani,2018_tanga,2021_osc}. Ideally, from such studies we might also find an observational signal which could serve to validate the hypothesis that neutrons decay into dark matter. This is the objective of our current work.

Here we focus on the consequences of ensuring that even if some neutrons convert to baryonic dark matter the total number of baryons in the star must be conserved. 
The brief layout of the article is as follows.
In section two, the equation of state (EoS) used for nuclear matter is presented, along with the Tolman-Oppenheimer-Volkov (TOV) equations~\cite{Tolman169,PhysRev.55.374,CIARCELLUTI201119,Sandin_2009} used to compute the properties of the neutron stars. In section three the consequences of the decay are explored and compared with the established constraints on neutron  star properties. This is followed in section four by a summary of our main results and conclusions.

\section{Neutron star matter}
A neutron star covers a huge range of energy densities right from the outer surface to the core. At the highest densities it must be said that we do not know with confidence in which form the matter exists at the core of the neutron star~\cite{Lawley:2006ps,Whittenbury:2013wma,Whittenbury:2015ziz,PhysRevD.4.1601,PhysRevD.30.272,Bombaci_2004,PhysRevD.102.083003,doi:10.1143/JPSJ.58.3555,2012_a,doi:10.1063/1.4909561,2016_a,PhysRevC.58.1804,BALBERG1997435,1985ApJ...293..470G,KAPLAN198657,PhysRevLett.79.1603,PhysRevLett.67.2414,Glendenning1997,Haensel2017,1980PhR....61...71S,weber2007neutron,2019_fri,Weber2016,Terazawa:2001gg,Husain_2021,2001_lattimer,2020_latti,2021_l,Cierniak:2021knt,Shahrbaf:2022upc,10.1143/PTP.108.703,2017_xyz,2022_xxyz,Motta:2022nlj}. Indeed it is the subject of active research. However, it is generally agreed that for the majority of neutron stars typically 90\% of the matter consists of neutrons, which makes neutron stars ideal candidates to test the hypothesis of neutron decay into DM. If neutrons do decay into dark matter a neutron star should contain enough dark matter particles to change the neutron star properties \cite{Mukhopadhyay_2017,PhysRevD.77.043515,PhysRevD.77.023006,CIARCELLUTI201119,Sandin_2009,Leung:2011zz,Ellis_2018,bell2020nucleon,2021_w,2000NuPhB.564..185M,Blinnikov:1983gh,2018_sa,2019_reddy,2013_red,berryman2022neutron,McKeen:2021jbh,deLavallaz:2010wp,Busoni:2021zoe,Sen:2021wev,Guha:2021njn}. To model the nuclear matter in a neutron star an equation of state based on the quark-meson coupling (QMC) model is adopted \cite{Guichon:1987jp,Guichon:1995ue,Stone:2016qmi,RIKOVSKASTONE2007341}. Here we have assumed that the hadronic content of the neutron star matter consists of nucleons only; that is, it does not contain hyperons or strange quark matter.

\subsection{QMC Model}
The QMC model is based on the MIT bag model~\cite{DeGrand:1975cf}, which treats baryons as bag of three relativistic quarks.  Interactions between these quark bags take place by meson exchange between the quarks in one baryon and those in another, with these quarks confined in color singlets. As the energy density increases the baryons experience large Lorentz scalar and vector mean fields, which in turn are expected to modify their internal quark structure~\cite{Guichon:2018uew}. Details of the QMC model can be found in Ref.~\cite{Guichon:2018uew}, with the most recent applications to finite nuclei in Refs.~\cite{Martinez:2020ctv,Martinez:2018xep}. A characteristic feature of this model is that the change in the internal structure acts to oppose the applied scalar field, giving rise to the scalar polarisability, $d$, as: 
\begin{equation}
M_N^*(\sigma) = M_N -  g_\sigma \sigma + 
\frac{d\tilde(g_\sigma \sigma)^2}{2} \, . 
\label{eq2.1}
\end{equation}
Here $M_N$ is the nucleon mass and $g_\sigma$ is the $\sigma$-nucleon coupling constant in free space calculated in the model. We expect the particular choice of hadronic EoS to make little difference in this study and choose the standard nucleon-only EoS presented by Motta {\it et al.} in Ref.~\cite{Motta:2019tjc}.

\subsection{Neutron decay inside the neutron star}
As explained in the introduction, we consider the possibility of neutron decay into DM through the process
 \begin{equation}
     n \xrightarrow \qquad \chi + \phi , 
 \end{equation}
 where $\phi$ is an extremely light dark boson, which will escape from the star, carrying away some energy. The baryonic DM particle, $\chi$, must lie within an MeV of the mass of the neutron and for practical purposes will be treated as degenerate with the neutron in the following calculations. The presence this dark matter will change the composition of the neutron star. In particular, it must be in chemical equilibrium with the neutrons. The $\beta$ equilibrium equations are given 
 as~\cite{Motta:2018bil,Motta:2018rxp}
\begin{equation}
    \mu_n = \mu_\chi \qquad
 \mu_n = \mu_p + \mu_e \qquad
 \mu_\mu = \mu_e \qquad
 n_p = n_e + n_\mu \, , \qquad
\end{equation}
where $\mu$ represents the chemical potential of each of the particles and $\mu_\mu$ represents the chemical potential of the muon. The Hartree term in the energy density of the system, including the dark matter, is given by:
\begin{equation}
\begin{aligned} 
    \epsilon_H = \frac{1}{2}m^2_\sigma\sigma^2 + \frac{1}{2}m^2_\omega\omega^2 + 
    \frac{1}{2}m^2_\rho\rho^2 + \frac{1}{\pi^2} \int^{k_F^n}_0 k^2 \sqrt{k^2 + M_N^*(\sigma)^2} dk + 
    \frac{1}{\pi^2} \int^{k_F^p}_0 k^2 \sqrt{k^2 + M_N^*(\sigma)^2} dk \\ +  \frac{1}{\pi^2}\int^{k_F^e}_0 k^2\sqrt{k^2 + m_e^2} dk + \frac{1}{\pi^2} \int^{k_F^\mu}_0 k^2 \sqrt{k^2 + m_\mu^2} dk + 
    \frac{1}{\pi^2}\int^{k_F^\chi}_0 k^2\sqrt{k^2 + m_\chi^2} dk 
    \, . 
\end{aligned}
\end{equation}

The full Fock terms, including the exchange of $\sigma, \omega, \rho$ and $\pi$ mesons, are given in Ref.~\cite{Motta:2019tjc}. The nucleon effective mass, $M_N^*(\sigma)$, was given in Eq.(~\ref{eq2.1}) and $k_F$ is the Fermi momentum of each particle species. The EoS used here has the normal nuclear matter density as 0.148 fm$^{-3}$, with binding energy per nucleon -15.8 MeV, symmetry energy 30 MeV, incompressibility 295 MeV and the slope of the symmetry energy 52.4 MeV. The pressure (P) can be calculated using
\begin{equation}
    P = \int \mu_f n_f - \epsilon.
\end{equation}
 The number of baryons are given by 
\begin{equation}
    B_i = 4\pi\int_0^R \frac{r^2 n_i(r)}{[1 - \frac{2M(r)}{r}]^{1/2}} dr,
\end{equation}
where i = $\chi$, p, n and n$_i$ is the baryon number density.

\subsection{Structure equations: Modified TOV equation for two fluids}
In order to satisfy the observational constraints, namely that neutron stars exist with masses as large as 2 $M_\odot$, it is necessary to allow the DM to be self-interacting. We take this interaction to be a simple vector interaction characterised by a strength parameter $G$ with the contribution to the energy density being $G n_{DM}^2/2$, with $n_{DM}$ the dark matter density. However, interactions with nuclear matter are ignored as they cannot be strong. Therefore, as the neutrons decay into dark matter, the neutron star will contain two different kinds of matter which do not interact with each other, the nuclear matter and the dark matter. For this reason, in modelling the properties of the neutron star, the two fluid~\cite{CIARCELLUTI201119,Sandin_2009} or modified TOV equations~\cite{Tolman169,PhysRev.55.374} are integrated with the EoSs.

The energy-momentum tensor for such a combination of fluids leads to the coupled equations:
\begin{equation}
\frac{dP_{nucl}}{dr} = - \frac{[\epsilon_{nucl}(r)+P_{nucl}(r)][4\pi r^3(P_{nucl}(r) +  P_{DM}(r))+m(r)]}{r^2(1-\frac{2m(r)}{r})},
\end{equation}
\begin{equation}
m_{nucl}(r) = 4\pi\int_{0}^{r}dr.r^2 \epsilon_{nucl}(r) ,
\end{equation}
\begin{equation}
\frac{dP_{DM}}{dr} = - \frac{[\epsilon_{DM}(r)+P_{DM}(r)][4\pi r^3(P_{nucl}(r) +  P_{DM}(r))+m(r)]}{r^2(1-\frac{2m(r)}{r})} ,
\end{equation}
\begin{equation}
m_{DM}(r) = 4\pi\int_{0}^{r}dr.r^2 \epsilon_{DM}(r),
\end{equation}
\begin{equation}
m(r) = m_{nucl}(r) + m_{DM}(r).
\end{equation}
Equations (2.4) to (2.8) have been solved by standard numerical methods, leading to the following results. In order to calculate the tidal deformability the method suggested by Hinderer {\em et al.}~\cite{Hinderer_2008,Hinderer_2010} has been adopted.

\section{Consequences of neutron decay into dark matter}
In this section the effects of neutron decay have been explored and they are tested against the known constraints on neutron star properties. We will focus on conservation of energy and total baryon number.
\begin{figure}[ht]
\centering 
\includegraphics[width=1\textwidth]{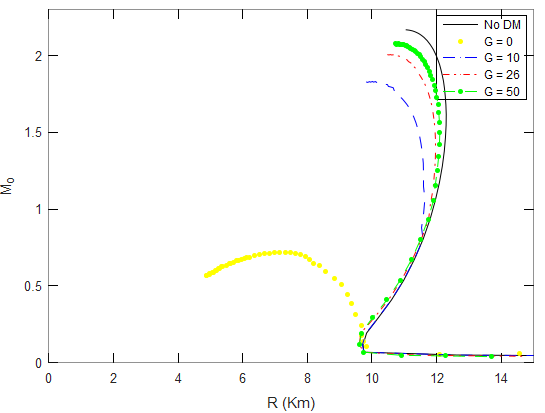}
\caption{Total mass versus radius of a neutron star for different strengths of the dark matter self interaction, G (in fm$^2$), at T = 0 $^\circ K$ before and after the neutrons decay.}
\label{fig1}
\end{figure}

\subsection{Mass and tidal deformability of the neutron star and the population of dark matter.}
Figure \ref{fig1} shows the relation between the mass and the radius of a neutron star. As the neutrons decay into dark matter, if the dark matter is non-self-interacting, the maximum mass of the neutron star falls from 2.23 M$_\odot$ to 0.7  M$_\odot$ \cite{Motta:2018bil}. To obtain a maximum mass of at least 2M$_\odot$, the dark matter repulsive vector self-interaction must be quite strong. Indeed, as shown in Fig.~\ref{fig1}, we find that in order to obtain a maximum neutron star mass of 2 M$_\odot$, the dark matter self-interaction has to have a strength given by G = 26 fm$^2$, which agrees with the results reported in  Ref.~\cite{Cline:2018ami}.
\begin{figure}[ht]
\centering 
\includegraphics[width=1\textwidth]{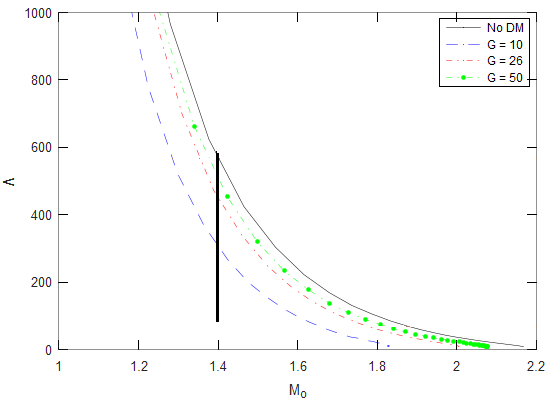}
\caption{Total mass vs tidal deformability of neutron star for  different dark matter self-interactions, G (in fm$^2$). The solid band indicates the constraint at around 1.4 $M_\odot$ from Ref.~\cite{Abbott_2017}.}
\label{fig2}
\end{figure}
A similar conclusion may be drawn from Fig.~\ref{fig2}, which shows the tidal deformability as a function of the mass of the star. From the gravitational wave measurement \cite{Abbott_2017,Abbott_2019,Bramante_2018_31} of the binary neutron star merger GW170817, the first analysis~\cite{Abbott_2017} suggested that a neutron star of mass 1.4 M$_\odot$ must have tidal deformability in the range, 70 $\leq \Lambda \leq$ 580, at 90\% confidence level. A later Bayesian analysis taking into account different priors corresponding to potential changes in the initial spin found a somewhat broader range~\cite{Abbott_2019}, as did the analysis of Zhao {\it et al.}~\cite{Zhao:2018nyf}.
From Fig.~\ref{fig2} it is evident that this constraint is satisfied even if neutrons decay into dark matter. 

\begin{figure}[ht]
\centering 
\includegraphics[width=1\textwidth]{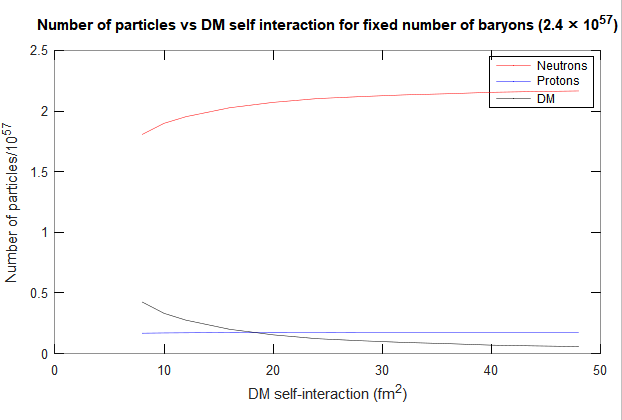}
\caption{Particle population against the strength of the DM self-interaction for neutron stars of different masses, when the total number of baryons is fixed at 2.4 $\times$ 10$^{57}$, corresponding to a star of mass before decay of 1.8 M$_\odot$. }
\label{fig3}
\end{figure}

In the process of neutron decay, described around Eq.~(\ref{eq:decay}), we do not expect the energetics to allow any baryonic matter to be lost. Thus, we expect that during the decay process the total number of baryons inside the neutron star must remain constant. The total energy of the system must not increase but we expect the escaping $\phi$ boson to carry away a significant amount of energy. 
Figure~\ref{fig3} shows the population of baryons as the strength of the DM self-interaction is varied. The relative population of DM particles decreases significantly as the DM self-interaction increases and, as the maximum mass constraint on neutron star requires DM to be strongly self-interacting,  the DM population must be quite low compared to the total number of neutrons that  it contains.

\subsection{Conservation of baryon number}
\begin{figure}[ht]
\centering 
\includegraphics[width=1\textwidth]{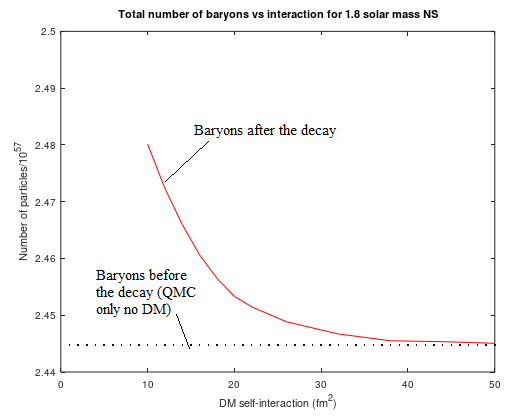}
\caption{Number of particles (baryons) vs DM self-interaction when keeping the total mass of the neutron star fixed at 1.8 M$_\odot$.}
\label{fig4}
\end{figure}
Figure~\ref{fig4}  shows the number of particles inside the neutron star as a function of the strength of the DM self-interaction for the case where the mass of the star is fixed. This naively indicates that the number of baryons inside the neutron star will not be conserved~\cite{PhysRevLett.121.061801}. Indeed, it suggests that if decay takes place, keeping the neutron star mass fixed, the number of particles inside the neutron star must increase. Clearly this is not possible and, as remarked earlier, we expect that the total number of baryons must be conserved. 

It is evident from Fig.~\ref{fig4} that for dark matter self-interactions at the lower end the increment in the number of particles is steep. We are interested in self-interactions with G $\geq$ 26 fm$^2$ (G in given in multiple of neutron-$\omega$ coupling) because below this value the neutron star does not follow maximum mass and tidal deformability constraints. At G = 26 fm$^2$ there would be approximately 2.5 $\times$ 10$^{55}$, or 1\%, more baryons inside the neutron star after the decay, were the mass to be held constant. 

\begin{figure}[ht]
\centering 
\includegraphics[width=1\textwidth]{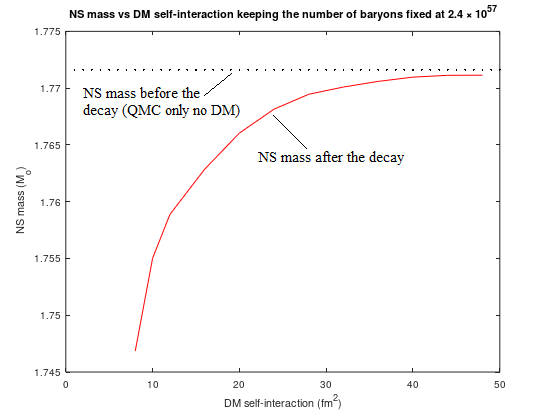}
\caption{Neutron star mass versus DM self-interaction strength (G in fm$^2$) when the total number of baryons inside the neutron star is fixed at 2.4 $\times$ 10$^{57}$. }
\label{fig5}
\end{figure}

\begin{figure}[ht]
\centering 
\includegraphics[width=1\textwidth]{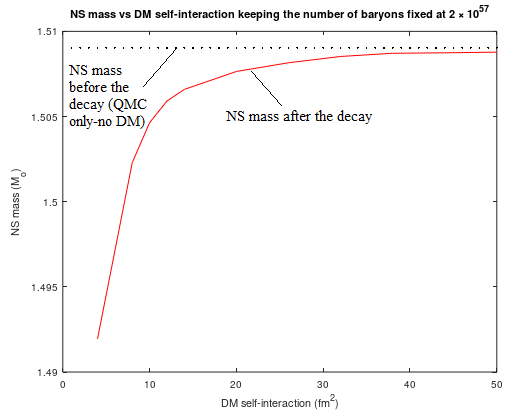}
\caption{Neutron star mass versus DM self-interaction strength (G in fm$^2$) when the total number of baryons inside the neutron star is fixed at 2 $\times$ 10$^{57}$.}
\label{fig6}
\end{figure}

To explore this issue further, in Figs.~\ref{fig5} and \ref{fig6} the mass of the neutron star is plotted against the strength of the self-interaction of the dark matter, while the total number of baryons inside the neutron star is kept constant (conserved) at 2.4 $\times$ 10$^{57}$ baryons and 2 $\times$ 10$^{57}$ baryons, respectively. In both figures the mass of the neutron star is not conserved when the decay process takes place. Indeed, they suggest that as the process of neutron decay takes place the neutron star mass decreases rapidly if the DM self-interaction is smaller than 26 fm$^2$ and the total number of baryons is conserved. In both figures,  at G = 26 fm$^2$ the change in the mass of the neutron star is approximately 0.002M$\odot$ or 0.14\%, which is quite large. Clearly this cannot happen unless we either emit this amount of energy, heat the star, or both.

\subsection{Change in temperature of the neutron star}
Neutron stars are such compact objects that their escape velocity is up to half the velocity of light. Within the scenario considered here the $\phi$ is extremely light and will be able to escape, carrying away significant energy. This will be especially so in the early stages of the decay process as neutrons near the top of the Fermi sea decay, releasing considerable energy; as the momenta involved are high the dark matter particles will tend to be deposited in states well above the lowest available. Thus after the decay the dark matter will not be a degenerate Fermi gas. 

It is interesting to explore the consequences for the star, given that  not all of the required decrease in mass  will be released as $\phi$ bosons. We note that the time scale for nuclear matter to cool below 1 MeV is of order seconds, far shorter than the time scale for neutron decay to dark matter (of order $10^5$ seconds). Thus we may treat the neutrons as a degenerate Fermi gas, while considering the effect on the properties of the star of the dark matter having a finite temperature. For the reasons explained in the previous paragraph, we expect the temperature of the dark matter to be considerably higher than that of the nuclear matter. If the neutron star heats up during the decay then it would naively be expected to expand, leading to a decrease in its rotational speed. On the other hand, as shown in Fig.~\ref{fig1} for a neutron star of fixed mass at a temperature of 0 $^\circ K$, the radius of neutron star is larger before the decay than afterwards. In order to resolve the question of whether the radius of the neutron star shrinks or expands following neutron decay, we compare the neutron star radius at T = $0^{\circ} K$ before the decay for a star containing nucleons only, to the radius of a star containing nucleons and DM with the same total baryon number and the DM heated by the energy equivalent of 0.001 M$_\odot$, which is a conservative upper limit (corresponding to the 1.4 M$_\odot$ case).
\begin{center}
	\begin{table}
		\begin{tabular}{|c|c|c|c|c|}
			\hline
		& NS Mass & T (MeV) &  R (Km) & I (kg.m$^2$)   \\
		\hline
		
		1 &	1.4 M$_\odot$ (Nuclear matter before decay) & 0  & 12.25 & 1.62 $\times$ 10$^{38}$ \\
			\hline
		2 &	1.4 M$_\odot$ (Nuclear matter + DM after decay) & 0  & 11.60 & 1.38$ \times$ 10$^{38}$    \\
		\hline
		\end{tabular}
		\caption{Properties of the  neutron star at temperature 0 MeV  before and after the neutron decay. Here, T stands for the temperature M represents the NS mass, R and I denotes the radius and the moment of inertia of the neutron star.}
		\label{Table:2}
	\end{table}
\end{center}
\begin{center}
	\begin{table}
		\begin{tabular}{|c|c|c|c|c|c|c|}
			\hline
		NS Mass & T (MeV)& $\Delta$ M & $\Delta$ R (m) & \% $\Delta$ R & $\Delta$ I (kg.m$^2$)  & \% $\Delta$ I\\
	
			\hline
			1.4 M$_\odot$ & 2 (DM) & 0.001 M$_\odot$  & 620 (decrease) &  5.06 \% & 2.28 $\times$ 10$^{37}$  & 14.1 \% \\		\hline
	
		\end{tabular}
		\caption{Changes in neutron star properties associated with a rise in the temperature of the dark matter following neutron decay. Here, T stands for the rise in temperature following the decay, $\Delta$M represents the equivalent changes in mass associated with the decay, $\Delta$R and $\Delta$I denote the change in radius (in meters) and the change in moment of inertia caused by the change in temperature. Here the radius of the neutron star after the decay shrinks compared to the radius of the neutron star before the decay. The changes in the values are given with respect to the first entry in Table ~\ref{Table:2}.}
		\label{Table:1}
	\end{table}
\end{center}

For a neutron star of mass 1.4 M$_\odot$ the rise in temperature corresponding to an energy equivalent of 0.001 M$_\odot$ would yield a temperature of 2 MeV, leading to a decrease in the radius by approximately 5.06 \%. A change of radius of this size significantly decreases the moment of inertia of the star, causing the star to spin up. The reduction in the moment of inertia is 2.28$\times$10$^{37}$ kg.m$^2$ or 14.1 \% when the radius decreases by 620 m, with the DM heated to 2 MeV. This would certainly lead to a substantial spin-up of the star over the time taken for the neutrons to decay.

That the radius of the star decreases even though the dark matter is at non-zero temperature may initially seem strange. However, we note that the decreases in radius shown in Tables~\ref{Table:2} and \ref{Table:1} are very close. That is, the temperature of the DM is more or less irrelevant, basically because, as shown in Fig.~\ref{fig3}, for suitably large values of the DM self-interaction the population of DM is less than 5 \% of the total baryon population.
%

\section{Conclusion}
As we have seen in Figs.~\ref{fig1} and \ref{fig2}, the requirement that neutron stars must be able to sustain masses as large as 2 $M_\odot$ requires that the dark matter formed in neutron decay must have a significant repulsive self-interaction. Indeed, assuming vector repulsion we find that the coupling strength must satisfy G $\geq$ 26 fm$^2$. With this self-interaction the stars also satisfy the constraint on tidal deformability deduced from the observation of GW170817. As a consequence of the strong repulsion we find that the population of dark matter inside the neutron star is much lower than that of normal matter.
Indeed, the lighter neutron stars have almost no dark matter when the dark matter self-repulsion G $\geq$ 26 fm$^2$. As shown in Fig.~\ref{fig3}, the relative population of dark matter particles is less than 5 \% even in heavier stars, with masses of order 1.8 M$_\odot$.

Even though the amount of dark matter formed is quite low, Fig.~\ref{fig4} indicates that the neutron star cannot conserve the number of baryons while keeping the total mass fixed.
Rather, as illustrated in Figs.~\ref{fig5} and \ref{fig6}, conservation of baryon number and conservation of energy mean that the star must heat up. 
While the small amount of heating of the nuclear matter will be radiated away  almost immediately by standard mechanisms, the DM will take much longer to cool. The primary mechanism will involve the collision of two DM particles yielding two neutrons which will not be entirely Pauli blocked because of the rapid cooling of the nuclear matter. This mechanism will be considerably slower than the rate of neutron decay to DM but eventually it will lead to a cold star. 

As shown in Table~\ref{Table:1}, the decay of neutrons to DM will lead to a significant  decrease in radius and hence moment of inertia. That in turn will lead to a sizeable increase in the rate of rotation of the star. While the dark matter will be formed at a finite temperature and only cool very slowly, we have seen that this has very little effect on the moment of inertia.
Given the mechanism for the decay, the time scale for the spin-up should be around 100,000 secs. This may in turn provide a signal for neutron decay if one is able to observe the neutron star sufficiently soon after its birth. \par

\acknowledgments
This research was supported by a University of Adelaide International Scholarship. It was also supported by the Australian Research Council through the Centre of Excellence for Dark Matter Particle Physics (CE200100008) and by the Alexander von Humboldt Foundation (TFM). 

\bibliography{main}
\bibliographystyle{ieeetr}

\end{document}